\documentclass[conference]{IEEEtran}
\usepackage[T1]{fontenc}
\IEEEoverridecommandlockouts
\usepackage{cite}
\usepackage{amsmath,amssymb,amsfonts}
\usepackage{amsthm}
\newtheorem{axiom}{Axiom}
\usepackage{algorithmic}
\usepackage{graphicx}
\usepackage{textcomp}
\usepackage{braket}
\usepackage{xcolor}
\def\BibTeX{{\rm B\kern-.05em{\sc i\kern-.025em b}\kern-.08em
    T\kern-.1667em\lower.7ex\hbox{E}\kern-.125emX}}
\begin{document}

\title{Comparing GHZ-Based Strategies for Multipartite Entanglement Distribution in 2D Repeater Networks\\
\thanks{MA thanks the graduate fellowship from the Pittsburgh Quantum Institute.}
}

%

\author{\IEEEauthorblockN{Mohadeseh Azari,$^*$ Amy Babay$^\S$,
Prashant Krishnamurthy$^\diamondsuit$, Kaushik Seshadreesan$^\dagger$}
\IEEEauthorblockA{\textit{Department of Informatics and Networked Systems, School of Computing and Information} \\
\textit{University of Pittsburgh}, Pittsburgh, PA, USA \\
$^*$moa125@pitt.edu, $^\S$babay@pitt.edu, $^\diamondsuit$prashk@pitt.edu, $^\dagger$kausesh@pitt.edu}
}

\maketitle

\begin{abstract} 
We conduct a comparative study to determine the initial quality necessary to extend the distance range of an $N$-qubit GHZ state (the parent state) using two-dimensional repeaters. We analyzed two strategies for distributing initial GHZ states using a centralized quantum switch to determine if any of the strategies show benefits: i) A strategy that employs quantum memories at the switch to retain quantum states entangled with each client node, where memory usage at the switch scales linearly with the number of clients, and ii) A strategy predicated on GHZ measurements at the switch node without reliance on memory assistance. In the former scenario, the switches generate GHZ states and teleport them to the clients by utilizing remote Bell pairs that are asynchronously generated and stored in memory. Conversely, in the latter scenario, the switches perform GHZ projective measurements on freshly generated remote Bell pairs without requiring local storage at the switch. 
To enhance the distance range of GHZ-type entanglement distribution, we analyze the two approaches as foundational elements for a self-repeating, two-dimensional quantum repeater architecture. 
Here, the clients of the switch nodes become the 2D repeater nodes that store elementary GHZ states in quantum memories, that can then be fused together to generate long-distance GHZ-type entanglement between end users of the network. By examining the two strategies' entanglement distribution rates and fidelities, we identify the conditions under which the 2D repeater architecture enhances overall performance, and we determine whether either method is a superior building block for such a repeater structure. Our findings illuminate the identification of effective modalities for the long-distance multipartite entanglement distribution within quantum networks.
\end{abstract}

\section{Introduction}
\subsection{Motivation}
Greenberger–Horne–Zeilinger (GHZ) type multipartite entangled states play a crucial role in numerous applications in quantum information processing. These applications include distributed quantum sensing~\cite{ho2024quantum}, fault-tolerant quantum computing~\cite{pankovich2023high}, clock synchronization~\cite{mckenzie2024clock}, conference-key agreement~\cite{wu2025semi}, and secret sharing~\cite{wang2024secure}. 
Due to their extensive applications, it is essential to identify the best method for distributing multipartite entangled states among end nodes in a quantum network. 
\subsection{Prior Works} 
Research to date has addressed direct and centralized switching architectures for short-distance GHZ distribution, the latter encompassing both memory-assisted~\cite{rozpedek2018optimal} and memory-free approaches~\cite {PhysRevA.111.022624} depending on hardware capabilities and design goals. While memory-assisted switching based on a central GHZ state source suffers from resource constraints, particularly in the presence of noise and memory decoherence, measurement-based switching schemes offer the potential for higher fidelity entanglement distribution ~\cite{massar2005classical, markham2008graph}. Centralized switching approaches can be further extended to incorporate quantum repeaters, enabling distance expansion through entanglement swapping and purification techniques~\cite{pant2019routing}. One class of repeaters, two-dimensional repeaters, is a great candidate for multipartite entangled states to extend the distance over which they are distributed. Despite these advancements, there remains an open question regarding which architectural choices for long-distance multipartite entanglement distribution best balance fidelity, rate, and scalability under realistic noise assumptions---a gap this work aims to address.
\subsection{Definition}
\begin{figure*}[t]
    \centering
    \includegraphics[width=0.9\textwidth]{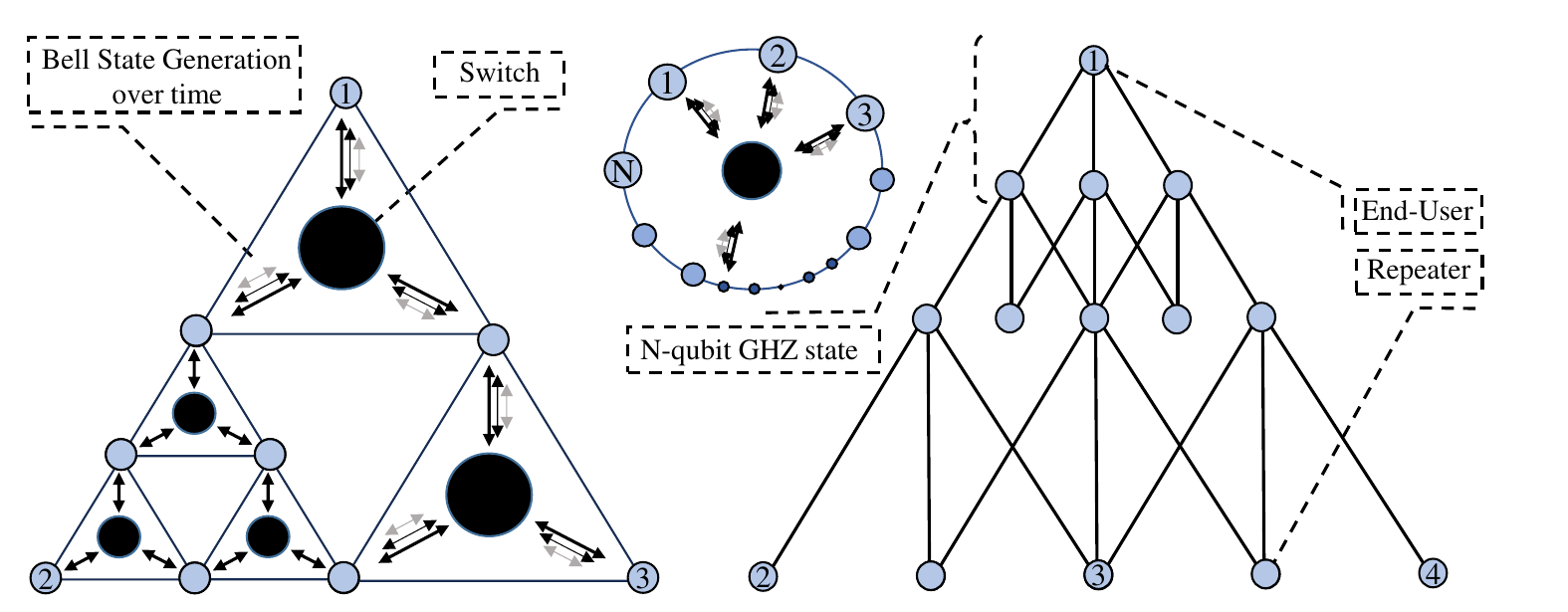}
    \caption{Schematic representation of distance extension using a two-dimensional repeater architecture. The left shows how three 3-qubit GHZ states, distributed via a centralized quantum switch, are fused to create a longer-range GHZ state spanning distant nodes. The right illustrates how this approach generalizes to any $N$ using centralized switching and Bell-state fusion.}
    \label{fig:GHZ_figure}
\end{figure*}
In this work, a quantum switch is defined as a node within a quantum network that enables the establishment of entanglement among its clients. A quantum switch functions as a versatile quantum device that may possess the capability to execute entangling gates, perform projective measurements (including Bell or GHZ-type measurements), and facilitate classical communication with other nodes. Depending on the specific switching protocol and the system's capabilities, such a node may or may not incorporate quantum memories as part of its operational framework.

We consider two primary switching paradigms. The first is a \emph{source-based} switch that generates a multipartite entangled state locally and distributes it to the end nodes. The second is a \emph{measurement-based} switch that performs a GHZ projective measurement on qubits entangled with remote users~\cite{PhysRevA.111.022624}. These switching strategies build on previous studies such as~\cite{rozpedek2018optimal}, which proposed centralized memory-assisted switch protocols for GHZ distribution, and~\cite{vivoli2015high}, a platform-aware analysis of practical implementations.
To extend the distance over which multipartite GHZ states can be distributed, quantum repeaters can work alongside quantum switches to complete the network~\cite{wallnofer2016two}. 

In this work, we recursively use a two-dimensional repeater structure to fuse short-range $N$-qubit GHZ states and construct a long-range $N$-qubit GHZ state among users who are physically more distant. Figure~\ref{fig:GHZ_figure} illustrates a two-dimensional repeater architecture. On the right, we show how 3-qubit GHZ states, each initially distributed by a centralized quantum switch, can be fused to create a single longer-range GHZ state. On the left, a graph representation demonstrates the same protocol for a 4-qubit GHZ state. For clarity and space considerations, the protocol for general $N$-qubit GHZ state distribution is not depicted; however, the same approach generalizes to arbitrary $N$, where the parent GHZ states can be created using centralized switch-based architectures and appropriate fusion operations~\cite{10622457}.

Our network of end nodes, switches, and repeaters engages three distinct types of nodes. The switch nodes collect Bell pairs from neighboring nodes and distribute GHZ entangled states to establish what the network calls a parent state. The repeater nodes fuse these parent states, extending the distance over which the GHZ is distributed. Lastly, the end nodes, representing a subgroup of the repeater nodes, are defined as the user nodes through which we intend to distribute the final GHZ state. While extending the distance, the 2D repeater protocol keeps the total number of qubits entangled in the GHZ state constant—$N$ initial GHZ states of $N$ qubits each are combined to produce a final GHZ state that still involves only $N$ qubits. As a result, each end node requires only a single qubit, while intermediate nodes may hold multiple qubits to facilitate entanglement fusion operations. This scheme reduces the overall impact of exponential photon loss across distance and enables entanglement distribution in large-scale networks.
\subsection{Research-Questions}
In this paper, we address the following two groups of research questions. 
\begin{itemize}
    \item The first group of questions compares two types of parent architectures. We aim to determine under what conditions, for given network parameters such as Bell-state measurement inefficiency, memory noise, and link generation success probability, the source-based switch architecture outperforms the measurement-based switch. This comparison will be conducted for the average distribution rate and fidelity.
    \item The second group of questions investigates whether a two-dimensional repeater architecture, built upon parent-level GHZ states, can enhance entanglement distribution performance. The two approaches used to generate these parent states—target-based and measurement-based switching—offer different trade-offs between rate and fidelity. The target-based switch utilizes memory-assisted nodes to distribute $N$ Bell pairs, thereby maximizing the distribution rate but at the cost of reduced fidelity due to memory noise. In contrast, the measurement-based switch prioritizes the fidelity of the final GHZ state by eliminating quantum memory, albeit at the expense of a lower average distribution rate.
     Given these trade-offs, we examine how the fidelity of the initially distributed parent states and the average rate of parent distribution affect the overall performance. We evaluate whether increasing the repeater depth ($m > 1$) yields improvements in both the final state fidelity and the distribution rate compared to the baseline case of direct distribution ($m = 1$).
\end{itemize}
\subsection{Roadmap}
The subsequent sections of this manuscript are structured as follows. In Section~\ref{sec:Architecture}, we present the overall architecture for long-distance distribution of multipartite entangled states. Since our approach relies on a self-repeating structure to extend the entanglement range across the network, it is essential to define foundational parent architectures that serve as building blocks for this scheme. We introduce two parent-level switching architectures: source-based and measurement-based—and then describe how these can be systematically integrated into a two-dimensional repeater framework to achieve scalable, long-range GHZ state distribution.
Section~\ref{sec:rate_analysis_parent} conducts a comparative analysis of the two parent architectures to identify the most suitable candidate for the two-dimensional architecture. This analysis pertains to rate and fidelity, elucidating the operational trade-offs between memory utilization and distribution efficacy. Section~\ref{sec:rate_fidelity_2D} utilizes the findings of the previous section to investigate whether the proposed two-dimensional architecture enhances distribution fidelity and rate under the optimal candidate of the parent architecture. Finally, in Section~\ref{sec:conclusion}, we summarize our findings and underscore several unresolved questions, including formulating a cost-function framework that could inform future architectural decisions within quantum networks.
\section{Architectures}
\label{sec:Architecture}
The distribution of multipartite entangled states (such as $N$-qubit GHZ states) among $N$ end nodes in a quantum network can be accomplished through two primary classes of architectures, depending on the distance between neighboring nodes. For relatively short distances, we analyze a group of architectures referred to as parent architectures (the reason will become clear soon). The parent architectures we examine here are based on a central quantum switch and include the \textit {source-based switch} and the \textit {measurement-based switch}. For greater distances, we explored a \textit {two-dimensional architecture} that can overcome the limitations of direct distribution.
\subsection{Parent Architectures}
Each end node connects to a central quantum switch via a quantum link in the parent architectures. After establishing Bell pairs between the switch and the end nodes, the central switch distributes an $N$-qubit GHZ state using one of two methods.
In the source-based switch architecture, once all Bell pairs are established between the switch and the end nodes, the switch prepares an $N$-qubit GHZ state. This prepared GHZ state is then teleported to the end nodes by performing a series of Bell-state measurements (BSMs) between the GHZ state and the corresponding qubits that share the Bell pairs with the end nodes. This architecture requires high-fidelity local GHZ preparation and robust two-qubit BSMs, benefiting from lower operational complexity during the distribution process.
In the measurement-based switch architecture, once Bell pairs are established between the switch and the end nodes, the switch performs an $N$-qubit GHZ projective measurement directly across its qubits. This measurement entangles the end nodes into an $N$-qubit GHZ state without explicitly preparing a source GHZ state.
Two methods can be taken to generate the Bell pairs between the switch and the end nodes. One way is that each end node attempts to distribute a Bell pair with the switch individually and, upon success, stores the qubit in a quantum memory while waiting for all other nodes to succeed. In this case, quantum memories are required, and memory noise must be considered as it affects the fidelity of the final state. In the second approach, all end nodes attempt Bell pair generation simultaneously until all succeed, with no intermediate storage. This eliminates the need for quantum memories and results in higher fidelity at the cost of a lower distribution rate due to simultaneous waiting. 
\subsection{Two-Dimensional Architectures}
\begin{figure*}[t]
    \centering
    \begin{minipage}{0.48\textwidth}
        \centering
        \includegraphics[width=\textwidth]
        {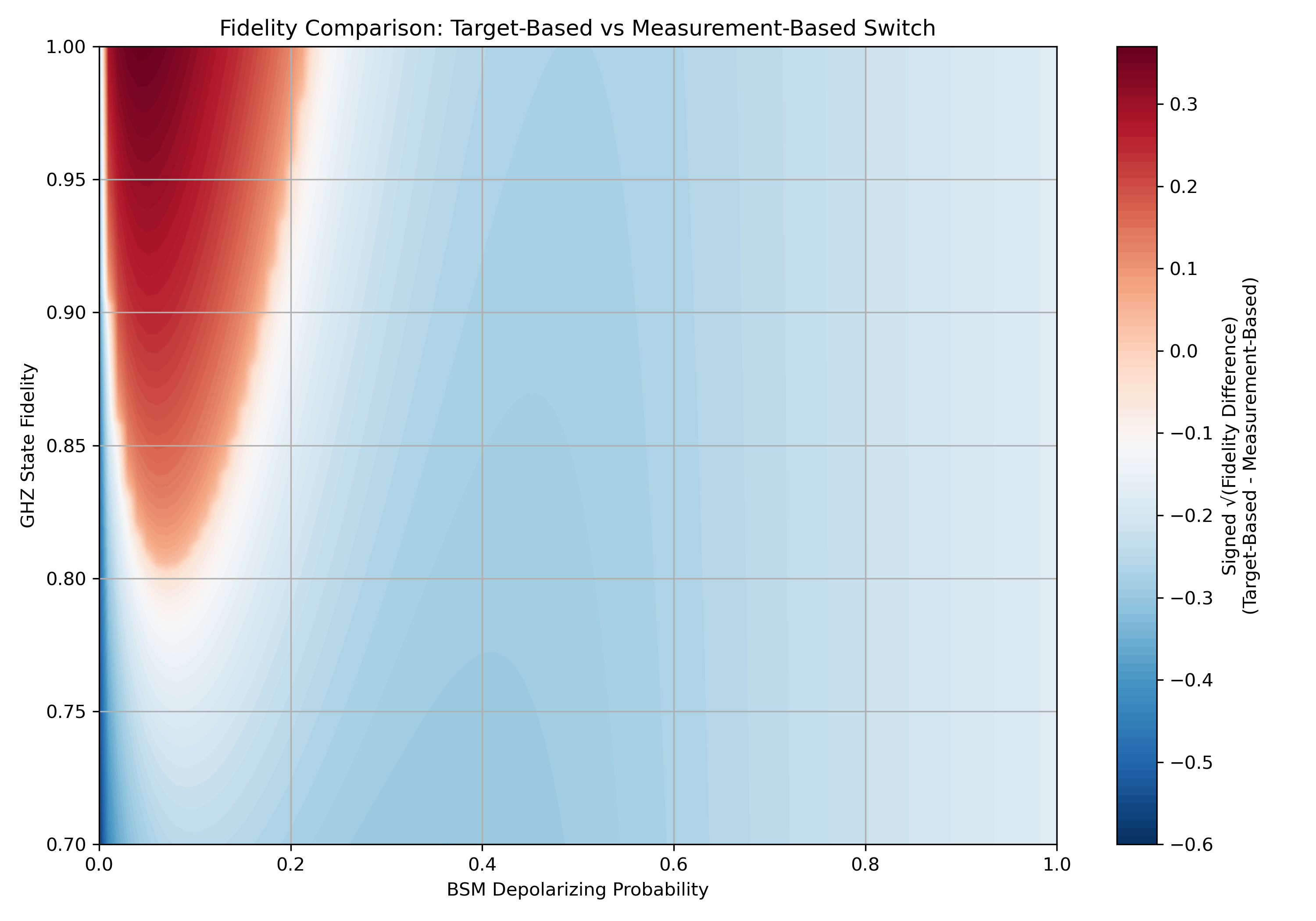}
        \caption{Fidelity difference between source-based and measurement-based switches as a function of BSM depolarizing probability and initial GHZ state fidelity. Positive values indicate regimes where the source-based switch performs better; negative values indicate measurement-based switch advantage. The fidelity advantage of the measurement-based approach becomes prominent at low GHZ fidelities and high BSM noise.}
        \label{fig:parent_fidelity_color}
    \end{minipage}
    \hfill
    \begin{minipage}{0.48\textwidth}
        \centering
        \includegraphics[width=\textwidth]
        {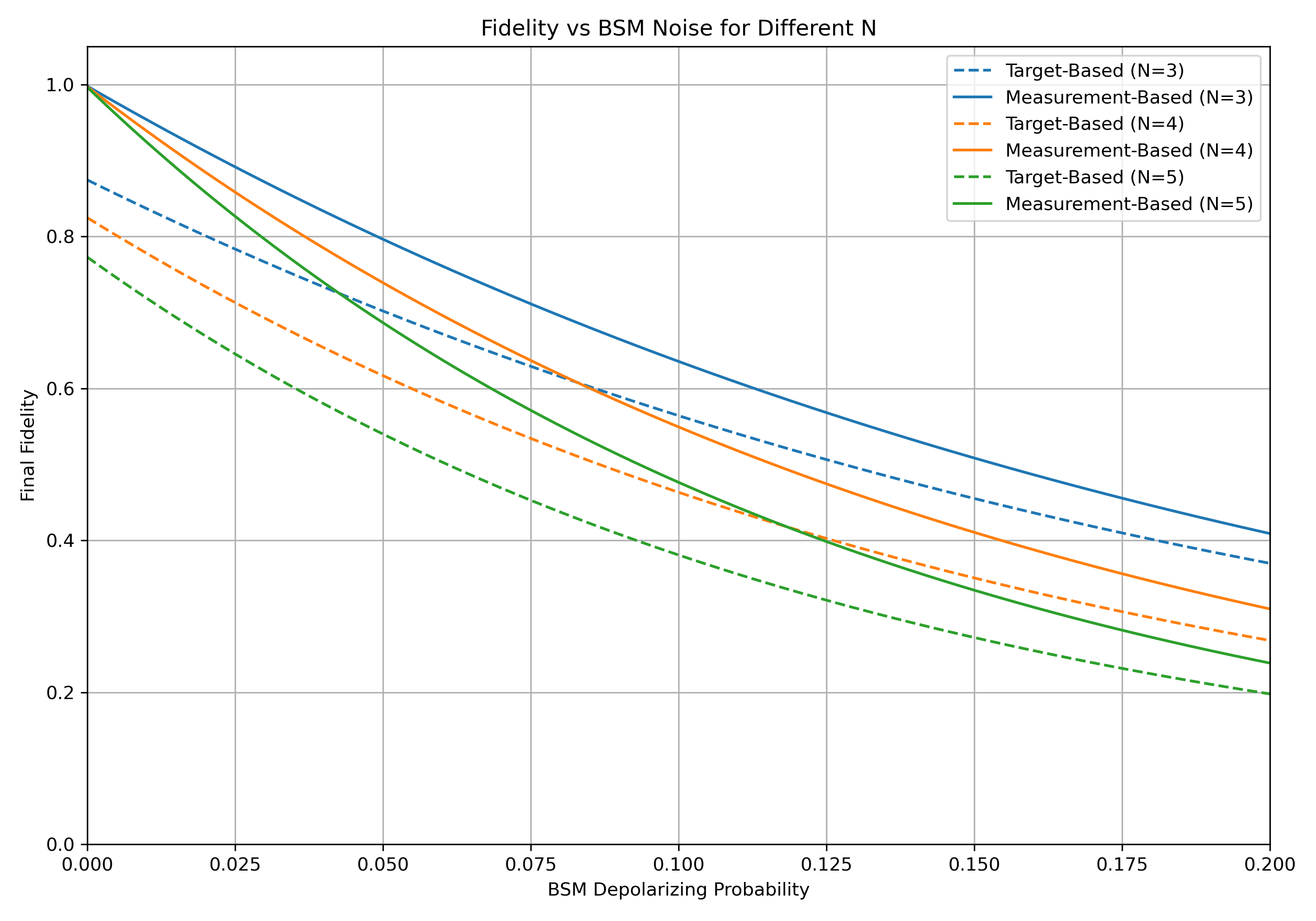}
        \caption{Final state fidelity as a function of BSM depolarizing probability for both source-based (dashed) and measurement-based (solid) switches across different numbers of users $N \in \{3,4,5\}$. The measurement-based switch maintains higher fidelity in all scenarios, especially as $N$ increases, highlighting its robustness to BSM noise due to the absence of memory decoherence.}
        \label{fig:parent_fidelity_line}
    \end{minipage}
\end{figure*}
Distributing directly becomes inefficient over greater distances between neighboring nodes due to the exponential decay of link success probability with distance. To address this issue, a two-dimensional repeater architecture can be utilized. This architecture first prepares elementary multipartite GHZ states across short distances. Then, multiple copies of these GHZ states are connected using local operations to create a GHZ state shared over a more considerable distance while keeping the number of qubits constant. We adhere to the first method outlined in the paper \cite{Wallnofer2016}.
The two-dimensional repeater approach operates in two stages. In the first stage, short-distance $N$-qubit GHZ states are prepared between neighboring nodes. In the second stage, groups of these elementary GHZ states are connected through types of Bell state measurements, leading to a long-distance GHZ state. The distance is effectively doubled at each repeater level while maintaining the number of end nodes. This process can be applied recursively to distribute $N$-qubit GHZ states over arbitrary distances.
By reducing the link distance through the initial preparation of GHZ states over short segments and using a repeater cycle that combines connection with optional multipartite entanglement purification, the two-dimensional repeater may enhance the probability of successful distribution, thereby improving the average rate and fidelity of GHZ-state distribution over long distances. We will discuss this in the following sections.
The choice between parent architectures and two-dimensional repeaters, as well as between source-based and measurement-based switches within the parent architectures, depends on specific network parameters, such as distance, memory capabilities at the nodes, and available hardware fidelities.
There exists a methodology by which an $N$-qubit GHZ state can be distributed by utilizing only three-qubit GHZ states as initial resources, thereby minimizing the required resources. This process necessitates the analysis of the cost function, which we will address in future research endeavors.
\section{Parent State Evaluation}
\label{sec:rate_analysis_parent}
\subsection{Fidelity}
In this section, we compare two switching architectures—source-based and measurement-based—in terms of fidelity. These two models differ significantly in their quantum memory use and operational assumptions.
In the \textbf{source-based architecture}, each end node establishes a Bell pair with a central switch and stores it in memory. Once all links are ready, the switch performs a sequence of Bell-state measurements (BSMs) to generate the final GHZ state. This model is memory-intensive and more demanding regarding hardware capabilities, as all entangled pairs must be stored until the final BSMs are executed. The fidelity in this model is affected by both memory depolarization noise and gate-level BSM noise.
In contrast, the \textbf{measurement-based architecture} avoids using memory altogether. In this case, each end node continuously attempts Bell pair generation with the switch until all succeed simultaneously. Only then are the GHZ measurements applied. As a result, memory noise is eliminated. The tradeoff, however, is that all link attempts must succeed in the same time step, which may reduce the overall rate of successful GHZ distribution. Nevertheless, the fidelity can be significantly improved due to the absence of decoherence from waiting in quantum memory.

Figure~\ref{fig:parent_fidelity_color} shows a contour plot comparing the fidelity difference between the source-based and measurement-based architectures. Positive regions (in red) indicate where source-based switching performs better, while negative areas (in blue) show where measurement-based switching is superior. The results suggest that the source-based approach can outperform the measurement-based one under low BSM noise. Still, as BSM depolarization increases, the measurement-based approach consistently yields higher fidelity. This reinforces the idea that the memory overhead in the source-based model degrades fidelity in practical scenarios.
Moreover, in Figure~\ref{fig:parent_fidelity_line}, we plot the absolute fidelity values for both architectures as a function of BSM noise for different values of $N$. The fidelity of the measurement-based switch can reach unity when BSM depolarizing noise is negligible, whereas the source-based switch saturates below one due to unavoidable memory-related degradation. As $N$ increases, the gap in fidelity becomes more prominent, emphasizing the scalability challenge of the source-based model.

While source-based switches offer a straightforward entanglement orchestration model, their reliance on quantum memory leads to fidelity bottlenecks. Measurement-based switches provide a path toward higher-fidelity GHZ state distribution, especially in settings where quantum memories are limited or unreliable. These findings motivate future exploration into hybrid architectures or advanced protocols that mitigate the probabilistic bottleneck of simultaneous link generation in measurement-based designs.
\subsection{Rate}
In the architecture under consideration, each end node must establish a successful Bell pair with the central switch before the switch can proceed with Bell state measurements. In the simplest scenario, each end node attempts to generate a single Bell pair at every time step. However, if the link success probability is low, the average number of time steps required to establish entanglement across all nodes becomes large, reducing the achievable rate.
To mitigate this issue, we consider allowing each end node to attempt \(m\) parallel Bell pair generations at each time step. An end node is considered successful if at least one of these \(m\) parallel attempts succeed. Mathematically, if the success probability of a single link attempt is denoted by \(q_{\text{link}}\), the effective success probability per node at each time step becomes
\begin{equation}
    q_{\text{eff}} = 1 - (1 - q_{\text{link}})^m.
\end{equation}
The corresponding average number of rounds required to successfully entangle all end nodes, denoted by \(\langle n_{\text{all}}(m) \rangle\), can be computed as
\begin{equation}
    \langle n_{\text{all}}(m) \rangle = \sum_{j=1}^{N} (-1)^{j+1} \binom{N}{j} \frac{1}{1 - (1 - q_{\text{eff}})^j}.
\end{equation}
Here, \(N\) is the number of end nodes. This expression accounts for the fact that all nodes must succeed before proceeding with the switching operation.
As \(m\) increases, \(q_{\text{eff}}\) approaches one, and the performance gradually improves toward the ideal case where each node succeeds with certainty at each time step. In the ideal limit as \(m \to \infty\), we recover the best possible performance, corresponding to
\begin{equation}
    \langle n_{\text{all}}(\infty) \rangle = 1.
\end{equation}
In the simulation, we analyze the quantity \(\langle n_{\text{all}}(m) \rangle - \langle n_{\text{all}}(\infty) \rangle\) as a function of \(m\). This quantity measures the gap between the performance at finite \(m\) and the ideal case. As expected, increasing \(m\) reduces the gap, but the improvement exhibits diminishing returns beyond a certain point. Therefore, infinite parallel attempts are not necessary in practice.
Instead, for a given performance threshold \(\epsilon\), we find the minimum value of \(m\) such that
\begin{equation}
    \langle n_{\text{all}}(m) \rangle - \langle n_{\text{all}}(\infty) \rangle \leq \epsilon.
\end{equation}
This optimization allows us to trade between the complexity of parallel attempts and the desired proximity to the ideal rate. We can ensure efficient operation without excessive resource overhead by carefully choosing \(\epsilon\).
\begin{figure*}[t]
    \centering
    \begin{minipage}{0.48\textwidth}
        \centering
        \includegraphics[width=\textwidth]{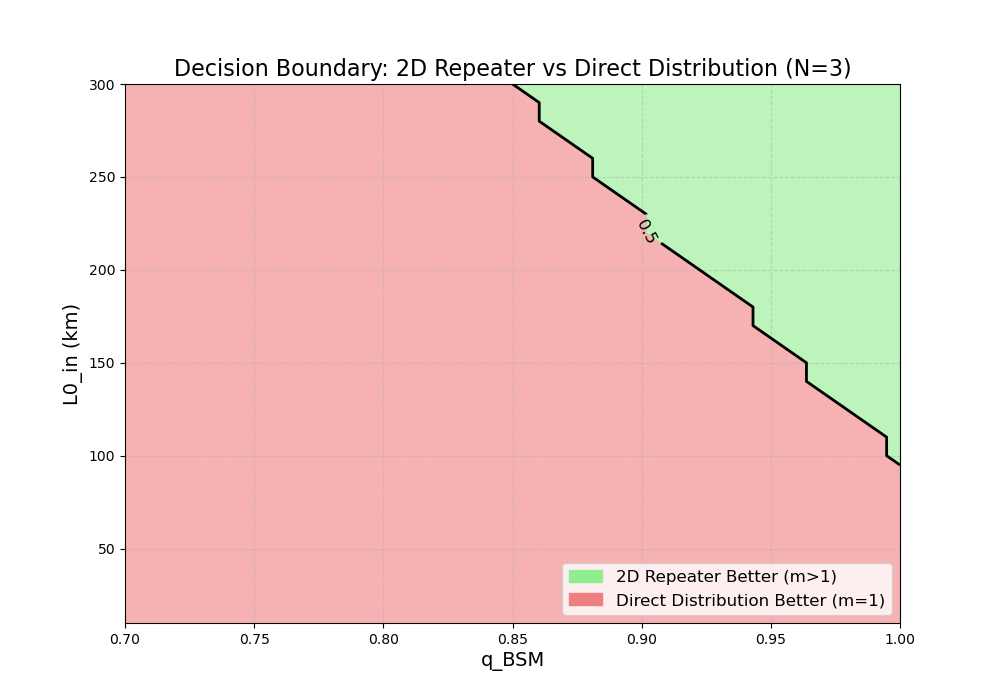}
        \caption{Decision boundary showing where two-dimensional repeater (\(m>1\)) outperforms direct distribution (\(m=1\)) in terms of average rate. Green regions indicate repeater advantage.}
        \label{fig:decision_boundary_rate}
    \end{minipage}
    \hfill
    \begin{minipage}{0.48\textwidth}
        \centering
        \includegraphics[width=\textwidth]{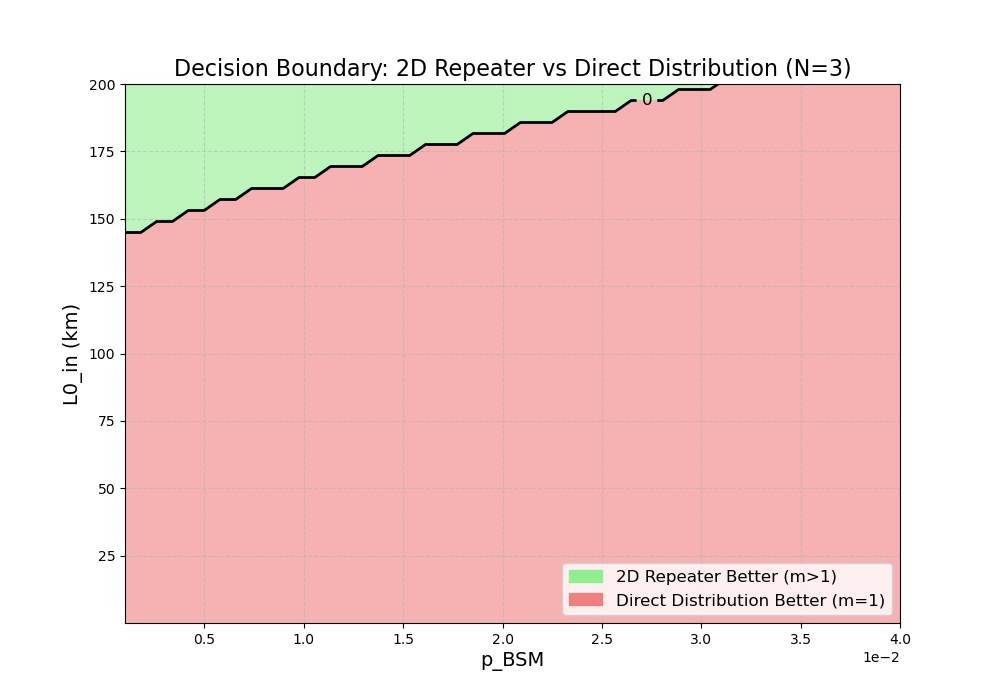}
        \caption{Decision boundary showing where two-dimensional repeater (\(m>1\)) outperforms direct distribution (\(m=1\)) in terms of final entangled fidelity. Green regions indicate repeater advantage.}
        \label{fig:decision_boundary_fidelity}
    \end{minipage}
\end{figure*}
\section{Distributed GHZ Evaluation}
\label{sec:rate_fidelity_2D}
After comparing the performance of the parent architectures in Section~\ref{sec:rate_analysis_parent}, namely the source-based switch and the measurement-based switch, we now extend the analysis to the two-dimensional repeater setting. In the two-dimensional repeater approach, a parent architecture is employed to distribute smaller-scale GHZ states over short-range links, and these elementary GHZ states are subsequently fused to extend entanglement over longer distances. By leveraging the source-based and measurement-based switches as the parent building blocks within the two-dimensional repeater, we aim to study how the rate and fidelity of GHZ-state distribution evolve with network parameters.
\subsection{Average Rate of Entanglement Distribution}
\subsubsection{Formulation}
This section explains how to derive the average rate of distributing an \(N\)-qubit GHZ state between \(N\)-symmetrically positioned nodes at the vertices of a polygon using a two-dimensional repeater approach. This section outlines the mathematical foundations for calculating the average number of time steps required to distribute an \(N\)-qubit GHZ state through an \(m\)-level two-dimensional repeater network. The methodology includes centralized and decentralized switching schemes for progenitor creation, with the analysis focused on quantifying the expected duration for successful link generation and final Bell state measurements (BSMs). 
\begin{equation}
    T = T_{\max} \times T_{\textrm{teleport}} \times \delta_t
\end{equation}
Let's denote \(T\) as the time required to distribute an \(N\)-qubit entangled state among \(N\) symmetrically allocated users through an \(m\)-level of 2D nesting. This time is determined by the time needed to generate \(N\) parents, represented as \(T_{\max}\), as well as the time required to perform Bell state measurements for distributing the parent's state to the child user's node, indicated by \(T_{\textrm{teleport}}\). Depending on the level of nesting (\(m\)), each parent is generated from its \(N\) parents. The key idea is to estimate the expected time \(E[T_{\max}]\) required for all necessary parent entangled states to be ready from the previous \(m-1\) nesting levels. Each child is produced from the \(N\) children of the last nesting level (now called parents), which means that at each level, every parent must wait until every one of the \(N\) parents has been successfully generated. This is why the time spent developing the parents is the maximum time for all parents. We can express the average time to generate \(N\) parents as the summation of its probability distribution. 
\begin{equation}
\begin{aligned}
    E[T_{\textrm{teleport}}] &= ({q_{\textrm{BSM}}}^{-1})^{N(N-1)/2},
    \\
    \quad E[T_{\max}] &= \sum_{k=1}^{\infty} \left[ 1 - F_{T}(k) \right].
\end{aligned}
\end{equation}
In the primary nesting level, whether using a centralized or decentralized method, an \(N\)-qubit multipartite entangled state, represented as an \(N\)-qubit GHZ state, is shared among \(N\) symmetrically arranged nodes. If the required level of nesting exceeds one, these nodes may not correspond to the end-user nodes. After distributing the \(N\)-qubit GHZ state across the \(N\)-symmetrically positioned nodes, we obtain \(N\) entangled nodes in the GHZ state, referred to as the initial parent because it represents the foremost nesting level. To determine the probability of establishing an initial parent in fewer than \(k\) time steps, represented as \(F_{T}(k,1)\), it is crucial to ascertain whether the initial parent was created using a centralized or decentralized approach. The following calculations will relate to both schemes. For systems with multiple repeater levels (\(m > 1\)), the overall CDF is defined recursively.
\begin{equation}
\begin{aligned}
    F_{T}&(k, m) = 
    \\
    &\left( \sum_{u=1}^{k} F_{T}\left(\lfloor k/u \rfloor, m-1\right) \times E[T_{\textrm{teleport}}] \right)^N
\end{aligned}
\end{equation}
\subsubsection{Analysis}
In this section, we analyze how the average rate of distributing an \(N\)-qubit GHZ state changes when employing a two-dimensional (2D) repeater architecture (\(m > 1\)) compared to direct distribution without repeaters (\(m = 1\)). The direct distribution protocol comprises two stages: (i) sharing Bell pairs between the central switch and each end node, and (ii) performing GHZ teleportation, either by preparing a source GHZ state followed by Bell-state measurements or through a direct GHZ projective measurement.
Without a two-dimensional repeater (\(m = 1\)), the protocol requires direct Bell-pair generation between the switch and each end node where nodes stand comparatively over a physical distance of \(L_0^{\text{in}}\) kilometers from each other. The entanglement generation success probability decreases exponentially with distance as:
\begin{equation}
    q_{\text{link}} = 0.5 \, \eta_c^2 \, e^{-L_0^{\text{in}}/L_{\text{att}}},
\end{equation}
where \(\eta_c\) is the coupling efficiency and \(L_{\text{att}}\) is the channel attenuation length. Consequently, the link generation process becomes the dominant bottleneck at large distances, reducing the GHZ distribution rate.
Introducing a two-dimensional repeater structure (\(m > 1\)) effectively divides the physical distance into shorter-range links. Bell pairs are created over these shorter links, and entanglement swapping is performed hierarchically. This structure significantly enhances the probability of link generation success, mitigating exponential decay with distance. However, it introduces additional Bell-state measurements at the switch, whose imperfections, characterized by the success probability \(q_{\textrm{BSM}}\), can accumulate and reduce the overall success rate.

\begin{figure*}[t]
    \centering
    \begin{minipage}{0.48\textwidth}
        \centering
        \includegraphics[width=\textwidth]{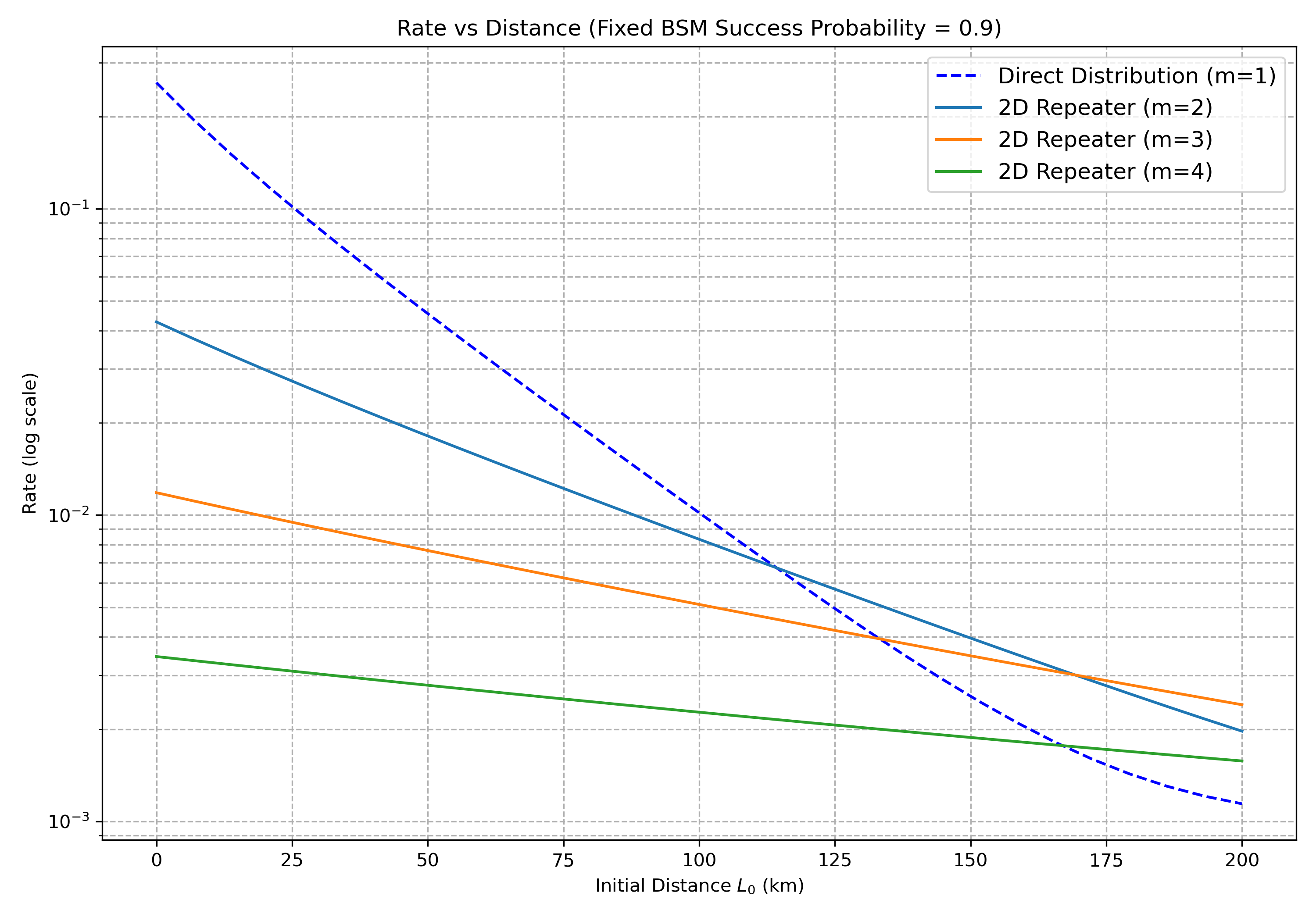}
        \caption{Average distribution rate as a function of distance (logarithmic scale). Comparison between direct distribution (\(m=1\)) and two-dimensional repeater architectures (\(m>1\)). The two-dimensional repeater approach enhances the achievable rate at larger distances.}
        \label{fig:2D_Rate_vs_distance}
    \end{minipage}
    \hfill
    \begin{minipage}{0.48\textwidth}
        \centering
        \includegraphics[width=\textwidth]{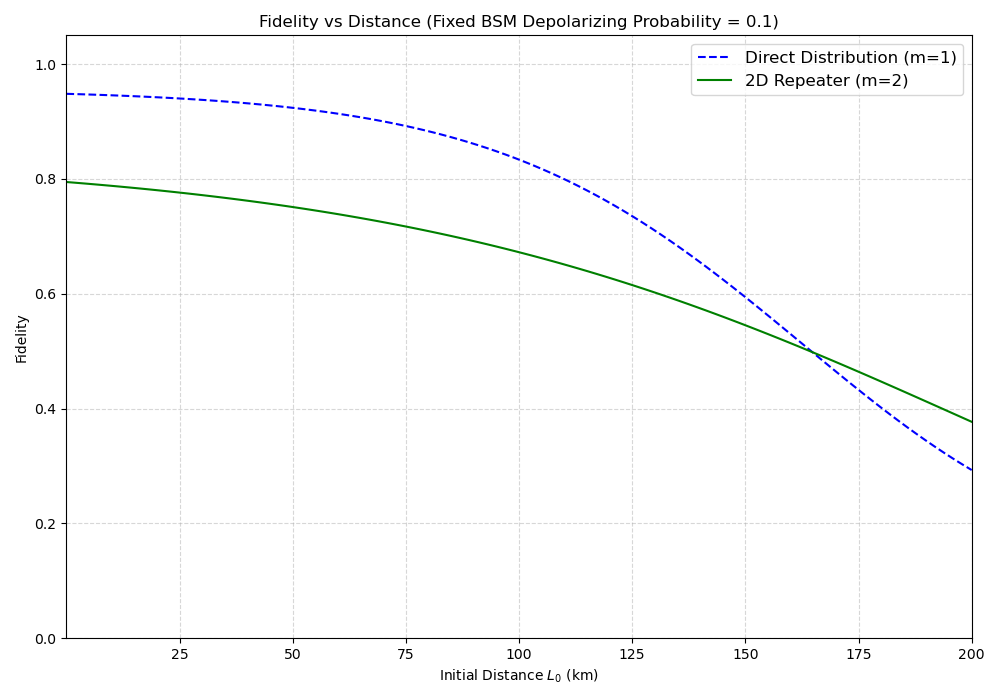}
        \caption{Fidelity of the distributed \(N\)-qubit GHZ state as a function of distance. Comparison between direct distribution (\(m=1\)) and two-dimensional repeater architecture (\(m=2\)). Switching to a two-dimensional repeater improves the fidelity at larger distances.}
        \label{fig:2D_Fidelity_vs_distance}
    \end{minipage}
\end{figure*}

Our numerical results, shown in Fig.~\ref{fig:decision_boundary_rate}, depict the parameter regimes where the two-dimensional repeater improves the GHZ distribution rate compared to direct distribution. It is observed that the two-dimensional repeater approach yields a higher rate when \(q_{\text{BSM}}\) is sufficiently high and/or when the distance between neighboring nodes \(L_0^{\text{in}}\) is large. Specifically, for fixed \(q_{\text{BSM}}\), increasing \(L_0^{\text{in}}\) shifts the advantage toward the repeater architecture because link generation becomes increasingly inefficient without subdivision.
Further insights are provided by Fig.~\ref{fig:2D_Rate_vs_distance}, where the GHZ distribution rate is plotted as a function of \(L_0^{\text{in}}\) for different nesting levels \(m\). For low \(L_0^{\text{in}}\), the direct distribution (\(m=1\)) outperforms the two-dimensional repeater due to the overhead of multiple Bell-state measurements. However, as \(L_0^{\text{in}}\) increases, the additional Bell-state measurements become less detrimental, and the repeater architecture (\(m>1\)) begins to surpass the direct distribution in rate.
In summary, the two-dimensional repeater improves the average GHZ distribution rate when the Bell-state measurement success probability \(q_{\text{BSM}}\) is sufficiently high and/or when the physical distance between end nodes is large enough. When \(q_{\text{BSM}}\) is low, direct distribution without repeaters remains favorable due to the reduced overhead.
\subsection{Average Fidelity of the Final State}
\subsubsection{Formalism}
We are considering a system consisting of $N$ parent states, where the state of the system is modeled as the tensor product of the individual parent states:
\begin{equation}
\sigma = \bigotimes_{i=1}^{N} \tau_i.
\end{equation}
Each parent $i$ is following:
\begin{equation}
\tau_i = p_{\text{GHZ}} {\ket{\text{GHZ}}\bra{\text{GHZ}}}_i + (1 - p_{\text{GHZ}}){\mathbb{W}_i^{\otimes N}}.
\end{equation}
As shown, with probability $p_{\text{GHZ}}$, parent $i$ is in an $N$-qubit GHZ state, and with probability $1 - p_{\text{GHZ}}$, parent $i$ is in the maximally mixed state shown as ${\mathbb{W}_i^{\otimes N}}$. Each parent consists of $N$ qubits, so that when we refer to GHZ states, we always mean an $N$-qubit GHZ state, and the maximally mixed state is between the $N$ qubits of parent $i$.
Thus, the state of the system can be expanded as:
\begin{equation}
\sigma = \sum_{k=0}^{N} \sum_{\substack{S \subseteq \{1,\dots,N\} \\ |S| = k}} 
\left({1 - p_{\text{GHZ}}}\right)^k \left(p_{\text{GHZ}}\right)^{N - k}
\bigotimes_{i=1}^{N} X_i,
\end{equation}
where
\begin{equation}
X_i = 
\begin{cases}
{\ket{\text{GHZ}}\bra{\text{GHZ}}}_i, & i \notin S \\
\mathbb{W}_i, & i \in S
\end{cases}.
\end{equation}
Again, going into the details of the above equation, we see that this represents the system's state as we open the tensor product. It describes three groups of possibilities: one possibility is that all the parents are in the GHZ state, which happens with probability $(p_{\text{GHZ}})^N$, meaning all the $N$ parents are in the GHZ state. The second possibility is when all the parents are in the maximally mixed state, which happens with probability $(1 - p_{\text{GHZ}})^N$, meaning all the parents are in the maximally mixed state. Finally, the third possibility is when $k$ number of parents are in the GHZ state and $N-k$ of them are in the identity, for which we have to choose that $k$ out of $N$.
It is important to emphasize that throughout this work, we consider single-qubit depolarizing noise only, and thus, the calculations presented are valid under this noise model.

After constructing the system's state, which represents the two-dimensional repeater architecture, the next step is to perform a series of Bell-state measurements. The purpose of these measurements is to distribute an $N$-qubit GHZ state between the end nodes, with the final GHZ state spanning almost twice the distance compared to the nodes in the parent system. The central question is how the system's state will change after applying the Bell-state measurements.
To understand this, we again look at three possibilities. The first possibility is when all the parents are in the GHZ state, which occurs with probability $(p_{\text{GHZ}})^N$. In this case, applying the Bell-state measurements will change the state of the system into a GHZ state between the final end nodes. The second possibility is when all the parents are in the maximally mixed state, which occurs with probability $(1-p_{\text{GHZ}})^N$. In this case, applying Bell-state measurements will leave the state of the system as the maximally mixed again. 
\begin{axiom}[Invariance of the Identity Under Bell Measurements]
Conducting Bell-state measurements on a system that commences in the maximally mixed (identity) state yields a final state that is once more the maximally mixed (identity) state.
\end{axiom}
The final possibility is when a group of $N-k$ parents is in the GHZ state and a group of $k$ parents is in the maximally mixed state. In this case, we need to pay a little more attention. There are three sub-possibilities:
If the Bell-state measurements occur only between the $k$ group of parents in the maximally mixed state, the final state will remain maximally mixed.
If the Bell-state measurements occur only between the $N-k$ group of parents in the GHZ state, the final state will remain a GHZ state.
Suppose a single Bell-state measurement exists between a qubit in the GHZ state and a qubit in the maximally mixed state. All quantum correlations are lost in that case, and the resulting state only has classical correlations.
\begin{axiom}[Destruction of Quantum Correlations]
The structure of Bell-state measurements in the two-dimensional repeater architecture is such that at least one Bell-state measurement connects any group of GHZ parents to any group of identity parents.
\end{axiom}
As a result, the only possibility of obtaining a final state with genuine quantum correlations is if all the parents are initially in the GHZ state. bellow is the state of the system after applying all the bell type measurements.
\begin{equation}
\begin{aligned}
\rho = \left(p_{\textrm{GHZ}}\right)^N & \zeta \left\{\ket{\text{GHZ}}\bra{\text{GHZ}}\right\} + \left(1 - p_{\textrm{GHZ}}\right)^N \mathbb{W}^{\otimes N} 
\\
+ \frac{1}{2} \sum_{\substack{U \subset \mathcal{N} \\ 1 < |U| < N}} 
&\left( \prod_{i \in U} p_{\textrm{GHZ}} \prod_{j \in \mathcal{N} \setminus U} 1 - p_{\textrm{GHZ}} \right) 
\\
& \zeta \left\{\mathcal{P}^{\otimes U} \otimes \mathbb{W}^{\otimes N \setminus U}\right\} .
\end{aligned}
\end{equation}
The $\zeta$ shows the single qubit depolarising channel acting on the final qubits. The channel parameters include the noise from the bell state measurements and the noise from the quantum memory since every parent should wait until all the parents are ready for teleportation.
\begin{equation}
\begin{aligned}
    G (N) = \bra{\text{GHZ}} \zeta & \left\{\ket{\text{GHZ}}\bra{\text{GHZ}}\right\} \ket{\text{GHZ}},
    \\
    \zeta \left\{\mathcal{P}^{\otimes U} \otimes \mathbb{W}^{\otimes N \setminus U}\right\} =& \sum_{\substack{U \subseteq \mathcal{N}}} 
    \left( \prod_{i \in U} \frac{1-p_i}{2} \prod_{j \in \mathcal{N} \setminus U} p_j \right).
\label{eq:G_func}
\end{aligned}
\end{equation}
Finally, the following is the fidelity of the final state that used repeated symmetric two-dimensional repeaters for distance expansion. The calculation of the function \ref{eq:G_func} might not be provided in the paper in favor of the limited space, but we will have detailed calculations in future work.
\begin{equation}
\begin{aligned}
{F} = \left(p_{\textrm{GHZ}}\right)^N &G(N) + (\frac{1}{2})^{N} \left(1 - p_{\textrm{GHZ}}\right)^N + \frac{1}{2}
\\
\sum_{\substack{U \subset \mathcal{N} \\ 1 < |U| < N}} 
&\left( \prod_{i \in U} p_{\textrm{GHZ}} \prod_{j \in \mathcal{N} \setminus U} \frac{1 - p_{\textrm{GHZ}}}{2} \right) 
\\
\sum_{\substack{V \subseteq \mathcal{N}}} 
&\left\langle \prod_{i \in V} \frac{1-p_i}{2} \prod_{j \in \mathcal{N} \setminus V} p_j \right\rangle,
\end{aligned}
\end{equation}
\subsubsection{Analysis}
In this section, we examine the behavior of the two-dimensional repeater architecture compared to direct distribution.
At small distances, the rate of generating quantum links is sufficiently high that the noise introduced by quantum memories is not a bottleneck. Although the two-dimensional repeater architecture improves the link generation rate in this regime, the extra noise added by the additional Bell-state measurements required in two-dimensional architectures overshadows this improvement. As a result, direct distribution (corresponding to $m=1$) remains advantageous for small distances. Only for considerable distances, where the link generation rate drops and memory noise becomes significant, does using the two-dimensional repeater become beneficial, as shown in Fig.~\ref{fig:2D_Fidelity_vs_distance}.
In Fig.~\ref {fig:decision_boundary_fidelity}, $p_{\textrm{BSM}}$ represents the depolarizing noise parameter associated with the Bell-state measurements. Higher values of $p_{\textrm{BSM}}$ introduce more noise into the system, diminishing the advantages of the two-dimensional repeater and shifting the threshold at which the two-dimensional repeater becomes beneficial to greater distances. Consequently, with increased Bell-state measurement noise, the system necessitates longer distances for the link generation bottleneck to surpass the noise penalty introduced by the two-dimensional repeater architecture.
\section{Conclusion}
\label{sec:conclusion}
In this study, we examined architectures for disseminating $N$-qubit GHZ states within a quantum network, initially concentrating on the foundational architectures—specifically, entanglement distribution methodologies utilizing a centralized quantum switch, which is capable of GHZ entanglement teleportation via either memory-assisted source or projective measurement.
The source-based switch, though technologically demanding, offers conceptual simplicity. It involves distributing Bell pairs between each end node and a central switch, followed by local Bell-state measurements at the switch. This approach is broadly compatible with existing quantum platforms but requires high-fidelity GHZ state preparation and quantum memories at the switch to store its share of Bell pairs. The memory requirement scales linearly with the number of end nodes ($O(N)$). In contrast, alternative architectures such as those in~\cite{vivoli2015high, patil2023distance} eliminate the need for switch-side quantum memory by modifying one protocol stage. These include either distributing multipartite entangled states directly between the switch and end nodes before performing asynchronous Bell-state measurements, or distributing Bell pairs as in the source-based approach and replacing the final step with GHZ projective measurements.

The memory-less architecture considered in this work corresponds to the measurement-based approach, in which all end nodes simultaneously attempt to establish Bell pairs with the central switch. Entanglement is generated only when all links succeed within a single time step. Although this strict synchronization requirement lowers the overall entanglement distribution rate, it significantly improves the fidelity of the resulting GHZ state. This is due to the absence of quantum memory, limiting the noise solely to link generation and measurement imperfections. When deploying a repeater-based architecture for distance extension, initiating the process with a high-fidelity state at the elementary level is critical, making memory-less, measurement-based switches attractive candidates despite their reduced success probability.

This rate limitation may be mitigated by parallel attempts or multiplexing Bell pair generation.
Finally, we explored two-dimensional repeater architectures and identified the conditions under which switching from direct GHZ state distribution to repeater-assisted architectures is beneficial. We examined the trade-offs in network performance across different parameter regimes through numerical simulations.

However, there is a way to increase the rate, which we will not analyze in this paper but will leave for future works. Each end node can distribute multiple copies of Bell pairs with the switch. By preparing various copies, one can balance the trade-off between rate and fidelity: having more copies available can increase the success probability of the overall distribution process and help maintain a higher final fidelity since redundant copies provide error resilience.
This strategy eliminates the need to wait during the link generation phase and enables a high-fidelity distribution of the GHZ state. Distributing multiple Bell pairs in parallel increases the overall rate while maintaining high fidelity and balancing operational complexity and performance. We leave the discussion of using error correction projective measurement on successful copies of the Bell pair for future works.

Our results raise several important open questions. Chief among them is the need for a platform-independent performance bound: for a given set of network parameters, what is the best achievable performance regardless of the specific architecture or physical implementation? Furthermore, defining a general and meaningful cost function that captures the trade-offs between fidelity, rate, resource consumption, and scalability remains an open and essential direction for future work.
\section*{Acknowledgment}

MA gratefully acknowledges support from Pittsburgh Quantum Institute (PQI) Graduate Fellowship. KPS acknowledges support from  PQI Community Collaboration Awards.

\bibliographystyle{IEEEtran}
\bibliography{Bib_File.bib}

\end{document}